\documentclass{aastex}
\usepackage{spr-astr-addons}
\usepackage{graphicx}
\usepackage{url}
\RequirePackage{color}
\newcommand{\emaila}{shntn05@gmail.com}

\begin{document}
\title{Do Pulsar Radio Fluxes violate the Inverse-Square Law?}
\shorttitle{Do Pulsar Radio Fluxes violate the Inverse-Square Law?}
\author{Shantanu Desai\altaffilmark{1}}
\email{\emaila}  
\altaffiltext{1}{Ronin Institute, Montclair, NJ 07043}

\begin{abstract} Singleton et al (2009) have  argued that the flux of  pulsars measured at 1400 MHz  shows an apparent violation of the inverse-square law with distance ($r$), and instead the flux  scales as  $1/r$. They deduced this from the fact that  the convergence error obtained in reconstructing the luminosity function of pulsars using an iterative maximum likelihood procedure is about  $10^5$ times larger for a distance exponent of two (corresponding to the inverse-square law) compared to an  exponent of one. When we applied the same  technique to this pulsar dataset with two different values for the trial luminosity function in the zeroth iteration, we find that neither of them can reproduce a value of $10^5$ for the ratio of the convergence error between these distance exponents. We then reconstruct the differential pulsar luminosity function using Lynden-Bell's $C^{-}$ method after positing both inverse-linear and  inverse-square scalings with distance. We show that this method cannot help in discerning between the two exponents. Finally, when we tried to estimate the power-law exponent  with a Bayesian regression procedure, we do not get a best-fit value of one for the distance exponent. The model residuals obtained from our fitting procedure  are larger for the inverse-linear law compared to the  inverse-square law. Moreover, the observed pulsar flux cannot be parameterized  only by  power-law functions of  distance, period, and period derivative. 
 Therefore, we conclude from our analysis using multiple methods that there is no evidence that the pulsar radio flux at 1400 MHz violates the inverse-square law or that the flux scales inversely with distance.
\end{abstract}

\keywords {pulsars; luminosity function; Malmquist bias; Lynden-Bell C-method }

\section{Introduction}

Pulsars are rotating magnetized neutron stars, which emit pulsed electromagnetic radiation. They have been detected throughout the electromagnetic spectrum. Pulsars are excellent laboratories for applications as well as tests of  nearly all branches of Physics \citep{Blandford}. Even though pulsars are mainly  located in our galaxy, they can also be used to address cosmological questions such as dark matter~\citep{Afshordi,Bramante,Desai} and constraining related modified theories of gravity~\citep{Freire}. At  the moment, 2524 pulsars are listed in the Australian Telescope National Facility  pulsar catalog~\citep{atnf}\footnote{http://www.atnf.csiro.au/research/pulsar/psrcat/expert.html}. In the last two decades there has been a plethora of new pulsar discoveries using  dedicated radio surveys and also from the {\it Fermi}~Large Area Telescope~\citep{fermi},  and one expects a ten-fold increase in the number of observed pulsars during the Square Kilometre Array era~\citep{Kramer}.

Radio telescopes typically measure the flux  of pulsars at multiple frequencies between 400-2000 MHz. Like most other astrophysical phenomenon, it is implicitly assumed that the pulsar fluxes obey the inverse-square law.  However, this {\it  ansatz} has been recently  challenged by~\citet{Singleton} (hereafter, S09). Their conclusions were  based on computing the convergence of a   maximum likelihood technique used to reconstruct the flux distribution (at 1400 MHz) of pulsars  from the Parkes multi-beam survey. If the results of S09 are correct, this would imply that either the distances to the pulsars are wrong by a factor of ten, or there is a component of the flux, which does not vary with distance ($r$) as $1/r^2$, thereby violating the inverse-square law. S09 argue that since the dispersion measure-inferred distances have been validated against other techniques, the most plausible conclusion is a violation of the inverse-square law. Such a violation  of the inverse-square law is expected in some proposed theoretical models of pulsar emission involving superluminal polarization currents~\citep{Ardavan}. This model has also been used  to explain the frequency spectrum of the Crab pulsar~\citep{Ardavan08}.

 However, the result of S09   has been disputed  by the pulsar community~\citep{Lorimer2010}. If the pulsar flux varied more slowly than an inverse-square law, one would have discovered a large number of pulsars in M31 or M33, whereas no confirmed detections have been made~\citep{Cordes03,Bhat,Rubio}. Nevertheless, it is possible that the non-detection of pulsars from these nearby galaxies could be due to other reasons such as: large pulse smearing or interstellar scattering along the line of sight; or the IMF and star formation rate in M31/M33  could be different from that in our galaxy; or accretion from dark matter  could  destroy the neutron star population~\citep{Bramante}, etc. Some of the above reasons have been invoked to explain the paucity of neutron stars in the Galactic Center~\citep{Dexter}.
  Therefore, given the potential path-breaking result claimed in S09, it is important to corroborate or refute their results with an independent analysis of the same dataset, which is the goal of this work. We check the claims of S09 using three independent methods to see if we reach the same conclusions.

The outline of this paper is as follows. In Section 2, we review the procedure followed by S09 to  test whether the pulsar fluxes obey the inverse-square law. In Section 3, we repeat the same procedure as S09, and describe in  detail our numerical methods. In Section 4, we reconstruct the pulsar luminosity function using  Lynden-Bell's $C^{-}$ method for two different distance exponents. In Section 5, we describe our parameter estimation method used to obtain the distance exponent. We conclude in Section 6.

\section{Review of S09 results}

S09 attempted to reconstruct the luminosity function of pulsars  detected from the Parkes multi-beam survey~\citep{Manchester}. The Parkes multi-beam survey  was the biggest ever pulsar survey carried out from 1997 to 2003 along the galactic plane with $|b|< 5^{\circ}$ and $l$ between $50^{\circ}$ and $260^{\circ}$. It was carried out with a 13-beam receiver  having a bandwidth of 288 MHz and a central frequency of 1374 MHz on the 64~m Parkes radio telescope. More details of this survey can be found in ~\citet{Manchester}.

For each pulsar, S09 calculated the distance from the observed dispersion measure using  the Cordes-Lazio model for the galactic distribution of free electrons~\citep{Lazio}, also known as  the NE2001 model in literature. S09 assumed that the luminosity function of pulsars is uniform throughout our galaxy and there are no population-specific selection effects. They demonstrated that the Parkes observations show evidence for Malmquist bias~\citep{Gonzalez}, since the cumulative flux distributions of the observed pulsars in different distance bins flattens out at  0.4~mJy. To circumvent this bias, they  applied the stepwise maximum likelihood method (SWML) ~\citep{Ellis}, which does not assume any functional form for the luminosity function  and corrects for the Malmquist bias. The SWML algorithm is routinely used in estimating the galaxy and quasar luminosity functions  from various extragalactic surveys~\citep{Ellis,Wilmer,Takeuchi}. 
We briefly recap the SMWL algorithm and discuss how it was applied to the Parkes dataset by S09. We use the same notation as  S09, which in turn followed the same notation as  ~\citet{Ellis}.

In SWML, the luminosity function $\phi(L)$ is determined nonparametrically  in  $N_b$ bins between a fiducial minimum and maximum value for the luminosity: $\phi(L)=\phi_k$, for $L_k-{\Delta L\over 2}<L<L_k+{\Delta L\over 2}$, with $k=1,...,N_b$, $\Delta L$ is the width of each luminosity bin, and $L_k$ is the luminosity in the $k^{th}$  bin.

The maximum likelihood  estimate for the luminosity function in each luminosity bin $k$ for a sample of  $N_p$  pulsars after the   $m^{th}$ iteration ($\phi_k^m$) is given by~\citep{Ellis}:
\begin{equation}
\phi_k^m \Delta L = \frac{\sum \limits_{i=1}^{N_p} W(L_i-L_k)}{\sum \limits_{i=1}^{N_p} \left \{ \frac{H[L_k-L_{\rm min}(r_i)]}{\sum \limits_{j=1}^{N_b}\phi_j^{m-1}\Delta LH[L_j-L_{\rm min}(r_i)]}\right \}}.
\label{eq:mlm}
\end{equation}

In Eq.~\ref{eq:mlm}, $L_{\rm min}(r_i)$ is the minimum detectable  luminosity for a pulsar at distance $r_i$, and  $\phi_j^{m-1}$ 
is the luminosity function after the  $(m-1)^{th}$ iteration in the $j^{th}$ luminosity bin.
  
 The window function $W(x)$ is given by
$W(x)=1$ for $ -\Delta L/2\leq x \leq \Delta L/2$, and 0 otherwise. We note that S09 report using 
the inverse of the above window function (See the sentence after Eqn~1 in S09).  $H(x)$ is given by 

\[
H(x)=\left\{\begin{array}{cl}
        0, & x \leq -\Delta L/2 \\
        (x/\Delta L + 1/2), & -\Delta L/2\leq x \leq \Delta L/2 \\
        1, & x \geq \Delta L/2.
           \end{array}\right.
\]

Therefore, the luminosity function can be obtained by iteratively  solving Eq.~\ref{eq:mlm} after
assuming an initial estimate for the luminosity function in each bin.

S09  applied the SWML algorithm  with an inverted window function (assuming no typographical error in their paper) to a sample of 1109 pulsars  from the Parkes multi-beam survey using the flux measured at 1400 MHz ($S_{1400}$). They parameterized the intrinsic luminosity, which is sometimes called pseudoluminosity in the pulsar literature~\citep{Bagchi} using $L=S_{1400} r^{n}$, where $n$ is the distance exponent, which quantifies how the flux scales with distance ($r$), and $S_{1400}$ is the measured flux at 1400 MHz. For each pulsar, they calculated the luminosity
from the observed flux and the putative power law index. They  solved Eq.~\ref{eq:mlm}  iteratively  in each bin to obtain the  final luminosity function after a fixed number of iterations. They parameterized the goodness of fit of their iterative procedure using a convergence  factor $\epsilon$, defined  as follows:
\begin{equation}
\epsilon = \sum \limits_{j=1}^{N_b}\left[\phi_m(j)-\phi_{m-1}(j)\right]^2,
\label{eq:epsilon}
\end{equation}
\noindent where the  difference in each luminosity bin ($j$) is between the $m^{th}$ and $(m-1)^{th}$ iteration, while solving Eq.~\ref{eq:mlm}. We note that the absolute value of $\epsilon$ depends on the normalization of the luminosity function.   S09 find that the luminosity function converges very rapidly for $n$=1.0 and 1.5,  and the convergence factor  is about $10^5$ times worse for the conventional inverse-square law ($n=2$). They obtain the best convergence for $n=1$ and $n=1.5$, depending on whether the full sample of pulsars  is analyzed ($n=1$), or if only  pulsars with periods less than 0.1 seconds are considered ($n=1.5$).  They applied the same procedure to a synthetic sample of simulated pulsars with both $n=2$ and $n=1.5$, and were able to demonstrate that they can  recover the original power law exponents.  Hence, they argued that their conclusions on the  violation of the inverse-square law are robust.

From this analysis, S09 conclude that if the distances to the pulsars are correct, the observed flux  at 1400 MHz falls off more slowly than $1/r^2$. However, they do not provide any information on the relative convergence error as a function of the number of iterations, or the number of luminosity bins used, or the trial luminosity  function in the zeroth iteration. Furthermore, it is entirely possible that the convergence factor for an inverse-square law could asymptote to the same value as $n=1$ (or 1.5) after increasing the number of iterations. Moreover, there is also no mathematical justification provided for using the above convergence factor as a metric for deciding on the best distance exponent.

\section{Application of the SWML Algorithm}

\begin{figure}
\begin{center}
\includegraphics[width=6cm]{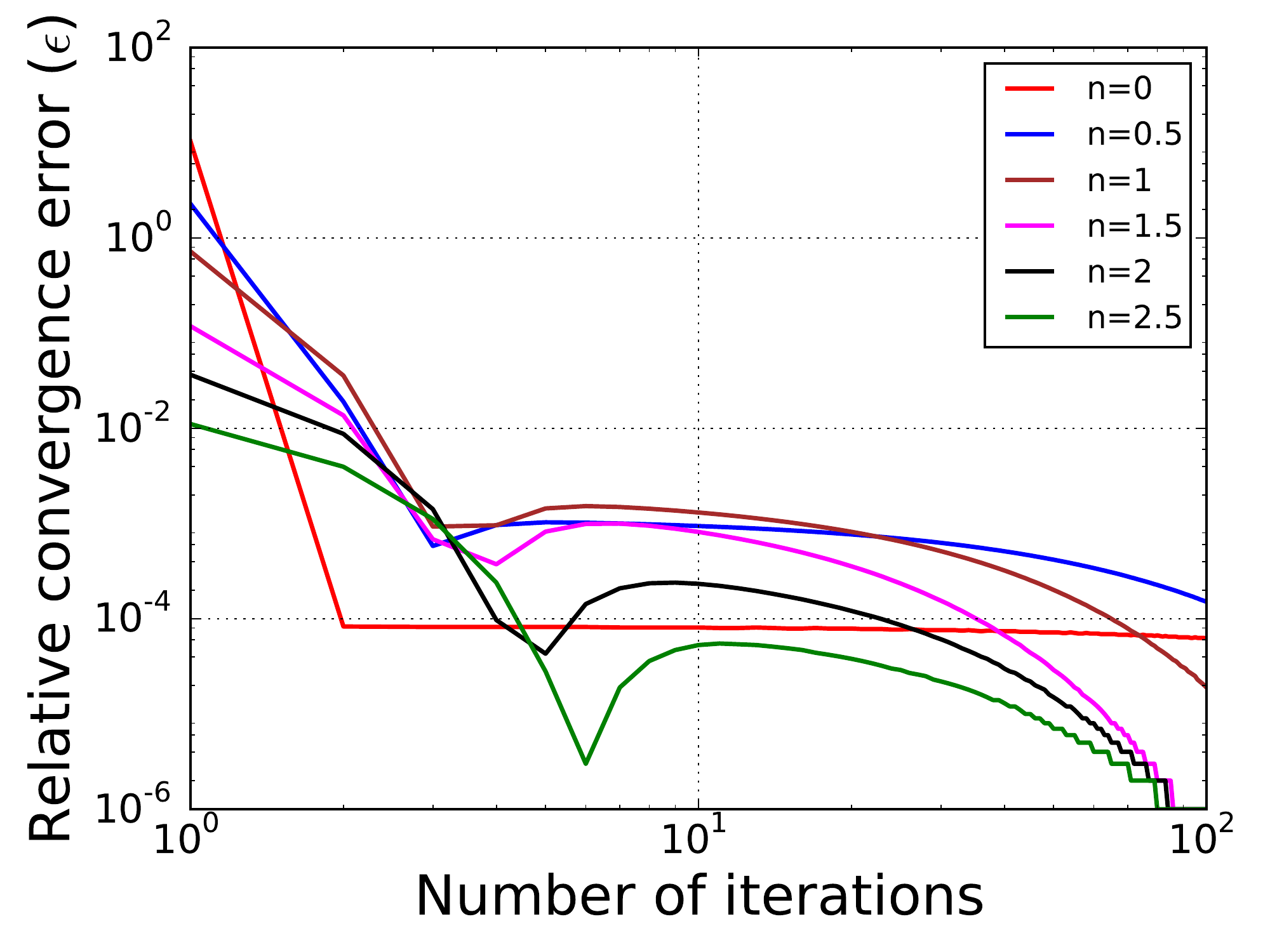}
\includegraphics[width=6cm]{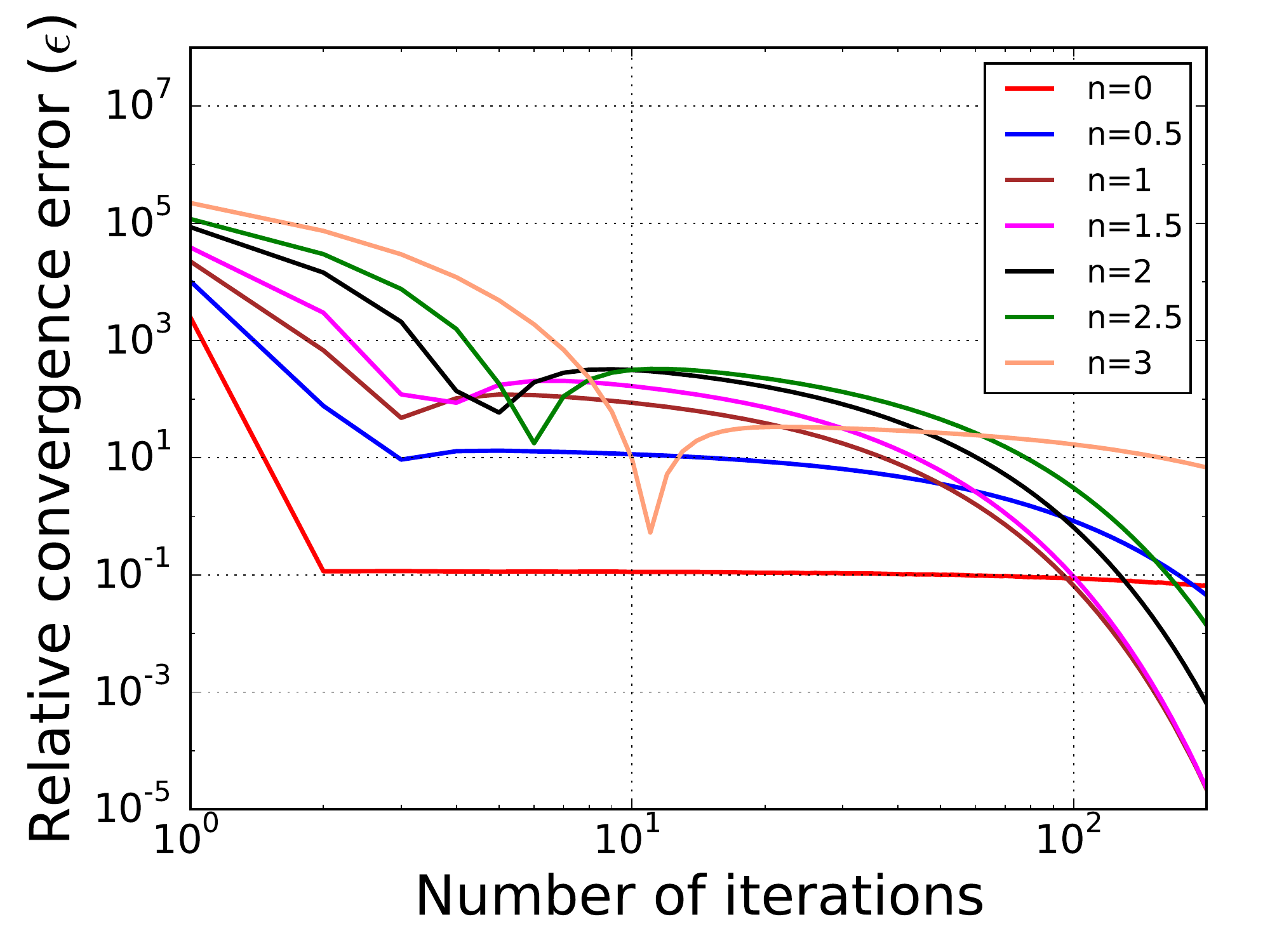}
\caption{Relative convergence error $\epsilon$ from the  SWML algorithm for pulsar fluxes  measured at 1400 MHz from the Parkes multi-beam  survey with two different luminosity functions assumed in the zeroth iteration.  The top panel shows 
the results for Method 1 and the bottom panel shows the same for Method 2. (See the caption of Table~\ref{tab:exponent} for explanation of both the methods). Therefore, we see no evidence that the convergence error for n=2 is five orders of magnitude larger than n=1 or 1.5.}
\label{fig:exponent}
\end{center}   
\end{figure}

Despite some of the concerns raised in the previous section regarding the conclusions of S09, we now try to replicate the same procedure as S09, to see if we can reproduce their results. We download the  Parkes multi-beam survey dataset from the ATNF online catalog. Similar to S09, we neglect  pulsars with $S_{1400} < 0.4$ mJy. We choose different putative values for the exponent $n$ ranging from zero to three in discrete steps of 0.5.  In order to use the SWML algorithm to self-consistently determine the luminosity function ($\phi$), one needs an initial starting value for the same  in different bins. Since no details are given in S09 about this, we apply the  SWML algorithm using two different guesses for the trial luminosity functions in the zeroth iteration. We denote these two applications of the SWML algorithm with different initial guesses  as Method 1 and Method 2.
In Method 1, we  assume that $\phi \propto L^{-1}$ for all values of $n$. This functional form
is similar to the empirically  derived luminosity function of pulsars~\citep{Bagchi}. In Method 2,
we use the  luminosities of pulsars (estimated from the observed flux and distance exponent) to construct the  luminosity function in each bin. This
procedure is similar to how  SWML  is used in extragalactic astronomy (C. Willmer, private communication). Therefore, in Method 1, the zeroth order luminosity function is the same for all distance exponents, whereas it depends on the power-law exponent in Method 2. In Eq.~\ref{eq:mlm}, $L_{\rm min}(r_i)$ for a pulsar located at distance $r_i$ is given by $0.4 r_i^n$ for a  particular exponent $n$. 
 Similar to S09, in both the methods, we also remove any outliers in luminosity for each exponent. 
For each power law exponent, we initially examine the luminosity distribution after splitting the data in about 50 bins, and then use the maximum luminosity bin with the smallest non-zero number of entries as the luminosity threshold.  We  then remove all pulsars with luminosity values exceeding this cutoff, so that there is at least one pulsar in every bin. After this outlier rejection, we then choose an optimum number of bins using   the Freedman-Diaconis algorithm, computed using {\tt astroML}~\citep{astroml}. According to Freedman-Diaconis rule, the optimal number of bins ($N_b$) for a dataset of $N$ points is given by~\citep{astroml}: 
\begin{equation}
N_b=\frac{2 (q_{75}-q_{25})}{N^{1/3}},
\end{equation}
where $q_{25}$ and $q_{75}$ are the first and third quartiles of the observed dataset.  
Since the SWML algorithm only provides information about the shape of the luminosity function and not its normalization~\citep{Wilmer}, we disregard the normalization between successive iterations. We note that the normalizations are different in Methods 1 and 2. We now discuss the results of applying the SWML algorithm for the different exponents.  The full set of luminosity thresholds, the number of pulsars after each cut, the number of bins, and the relative convergence error after 100 and 200  iterations  for Methods 1 and 2 are shown in Table~\ref{tab:exponent} for different power law indices.  A graphical summary of our results using both the  methods can be found in Fig.~\ref{fig:exponent}. We highlight the key results from both the methods below:

\begin{itemize}
\item {\bf Method 1}:
The values of $\epsilon$ after 100 iterations using Method 1 can be found in Table~\ref{tab:exponent}, and as a function of the number of iterations in the top panel of  Fig.~\ref{fig:exponent}. We note  that  the SWML algorithm with Method 1 fails  for $n=3$, since the luminosity function asymptotes to zero in all  the bins after 100   iterations. 
 We find that $\epsilon$ after 100 iterations for $n=2$ has the same value as that for $n=1.5$ and $2.5$, of  about $10^{-6}$. The value for $n=1$ is about ten times larger than for $n=2$. Therefore,  $\epsilon$ is of  the same order of magnitude for most  exponents.  The slope in  Fig.~\ref{fig:exponent} is also  the  same for $n$ between 1.5 and 2.5.  Therefore, the values we get for the ratio of  $\epsilon$ for $n=2$ compared to $n=1$ disagree with  those of S09.

\item {\bf Method 2}:
The values of $\epsilon$ for Method 2 after 200 iterations is shown in Table~\ref{tab:exponent}, and as  a function of the number of iterations in the bottom panel of Fig.~\ref{fig:exponent}. Unlike Method 1, the final luminosity functions  for each exponent do not converge after 100 iterations, and hence we doubled the number of iterations compared to Method 1, before examining the relative values of $\epsilon$ for different power law exponents.  
Using this method, we find that for $n=2$, $\epsilon$ is only about a factor of ten larger than
the same for $n=1$ or $n=1.5$. Therefore, although we agree with S09, that $\epsilon$ after a certain number of iterations  is smaller for $n=1$ and 1.5 as compared to $n=2$, the ratio is only about a factor of ten and not $\mathcal{O}(10^5)$ as claimed by S09. The slope is also the same between these three  exponents. 

\end{itemize}

\begin{figure}[t]
\centering
\includegraphics[width=7cm]{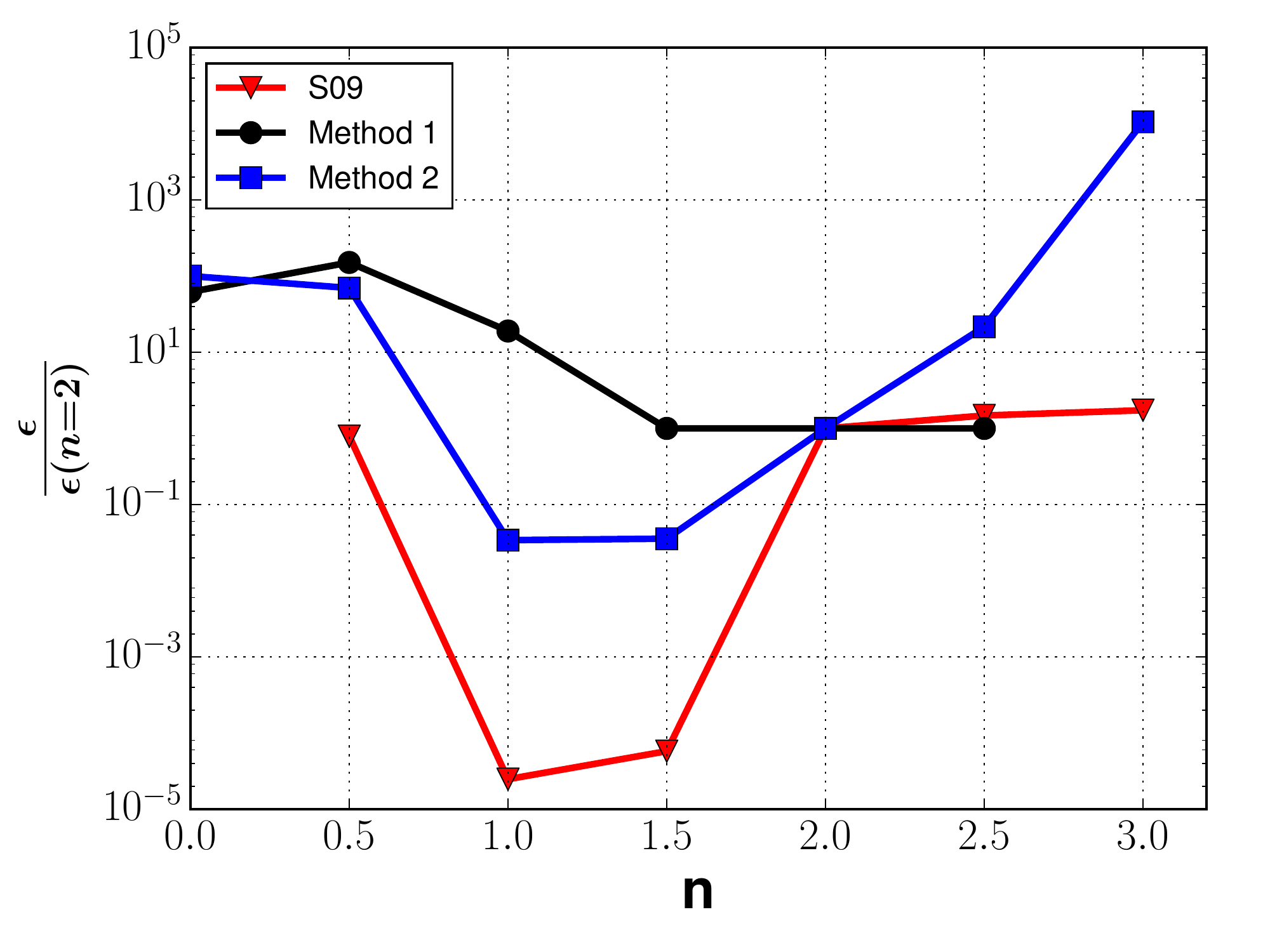}
\caption{Comparison of the convergence error ($\epsilon$) as a function of the distance exponent ($n$) normalized to its value at $n=2$ using our application of the SWML algorithm with  two different guesses for the initial 
luminosity function (Method 1 and 2) vs  the same for S09 (data obtained from Fig. 2 of~\citet{Singleton}). The value of $\epsilon$ for $n=2$ is approximately the same as $n=1$ for Method 1 and about ten times larger in Method 2. This does not agree with the results from S09, who find that $\epsilon$ for $n=2$ is about $10^5$ times larger than for $n=1$.}
\label{fig:comparison}
\end{figure}

Therefore, in summary we conclude that the convergence error of the SWML algorithm is sensitive to the 
initial guess for the trial luminosity function, before the iterative procedure is started. We have used  two different choices for these. If we assume that $\phi \propto 1/L$, then $\epsilon$ for $n=2$ is of the same order of magnitude as $n=1$ or $n=1.5$. On the other hand, if we use the observed data to construct the zeroth order luminosity function, the convergence error is larger for $n=2$ compared to $n=1$ and 1.5, by only a factor of ten. A comparison of the relative convergence error using both the guesses for the initial luminosity along with the same  obtained by  S09 is shown in Fig.~\ref{fig:comparison}. Note that all the three curves in Fig.~\ref{fig:comparison} have been normalized to the value of  $\epsilon$  at $n=2$, in order to compare the relative convergence errors for $n=1$ or 1.5. Therefore, we do not concur with  S09 after replicating their procedure. One possible reason for the large convergence error found by S09  for $n=2$, compared to $n=1$ or 1.5 could be due to the number of bins they  may have used  or their initial guess for the luminosity function. 

Another possibility could due to the inverted value for the window function  used by S09 compared to what is prescribed in the SWML algorithm.  To test this, we then applied the SWML algorithm with the same inverted window function. However, with this inverted window function, the SWML algorithm does not converge for either of the trial functions used in the zeroth iteration, and the luminosity function diverges to $NaN$ in all the bins. This implies that there is a typographical error in the reported window function in S09, and they used the same window function as in the original SWML paper~\citep{Ellis}.

\begin{table*}
\small
\caption{Results from the application of SWML algorithm to the flux of pulsars measured at 1400~MHz with different power-law distance exponents   using two different guesses for the luminosity function ($\phi$) in the zeroth iteration. In Method 1, we assume that $\phi \propto 1/L$, where $L$ is the luminosity in each bin. In Method 2, we use the luminosities of the observed pulsars to construct an empirical luminosity function. $L_{max}$ is the luminosity cut for each trial exponent and  $\epsilon$ is the convergence error defined in Eq.~\ref{eq:epsilon}. We also find that Method 1 is unable to reconstruct the pulsar luminosity function for  $n=3$.}
\label{tab:exponent}
\begin{tabular}{@{}lccccc@{}}
\tableline
 Exponent & $L_{max}$ & \# Pulsars & \# Bins  & $\epsilon$ (Method 1)  & $\epsilon$ (Method 2) \\
  & & &  & 100 iterations & 200 iterations \\ 
\tableline
0 & 5 mJy & 687 & 30 & $6.2 \times 10^{-5}$ & $6.4 \times 10^{-2}$ \\
0.5 & 15 mJy $\mathrm{kpc^{0.5}}$ & 691  & 40 & $1.5 \times 10^{-4}$ & $4.5 \times 10^{-2}$ \\
1.0 & 40 mJy kpc & 692 & 40 & $1.9 \times 10^{-5}$ & $2.2 \times 10^{-5}$ \\
1.5 & 100 mJy $\mathrm{kpc^{1.5}}$ & 687 & 30 & $10^{-6}$ & $2.3 \times 10^{-5}$  \\
2.0 & 200 mJy $\mathrm{kpc^2}$ & 677  & 25 & $10^{-6}$ & $6.4 \times10^{-4}$ \\
2.5 & 600 mJy $\mathrm{kpc^{2.5}}$ & 677 &  30 & $10^{-6}$ & $1.4 \times 10^{-2}$ \\
3.0 & 1400 mJy $\mathrm{kpc^3}$ & 663 & 22 & - & $6.8 $  \\
\tableline
\end{tabular}
\end{table*}

\section{Application of Lynden-Bell $C^{-}$ method}
All estimates of the cumulative luminosity function of pulsars in literature have been obtained after positing an inverse-square law~\citep{Bagchi}. In all these cases, the empirically derived luminosity function agrees well with the observed distribution and there have been no concerns about the mismatch between the data and the reconstructed luminosities. However, most empirical methods of estimating the pulsar luminosity function in literature  do not account for  the Malmquist bias in the observed flux distribution. So, we would like to apply a different  maximum likelihood technique, which is similar in spirit to SWML with two different power indices to see if the reconstructed luminosity functions in both the cases help distinguish between the two scenarios.
The algorithm we apply for this purpose is the  Lynden-Bell $C^{-}$ method~\citep{Lynden-Bell}, which was originally used to estimate the luminosities of quasars. Similar to SWML, $C^{-}$  is a non-parametric method to correct for the truncated distribution  and makes no assumption about the functional form for the luminosity. 
 As pointed out by ~\citet{Wilmer}, $C^{-}$ is the limiting case of SWML, where each bin contains only one object.  
However, it is not an iterative procedure like SWML.  We briefly recap the usage of the  binned version of the $C^{-}$ method. More details can be found in ~\citet{book}.

\begin{figure}
\begin{center}
\includegraphics[width=6cm]{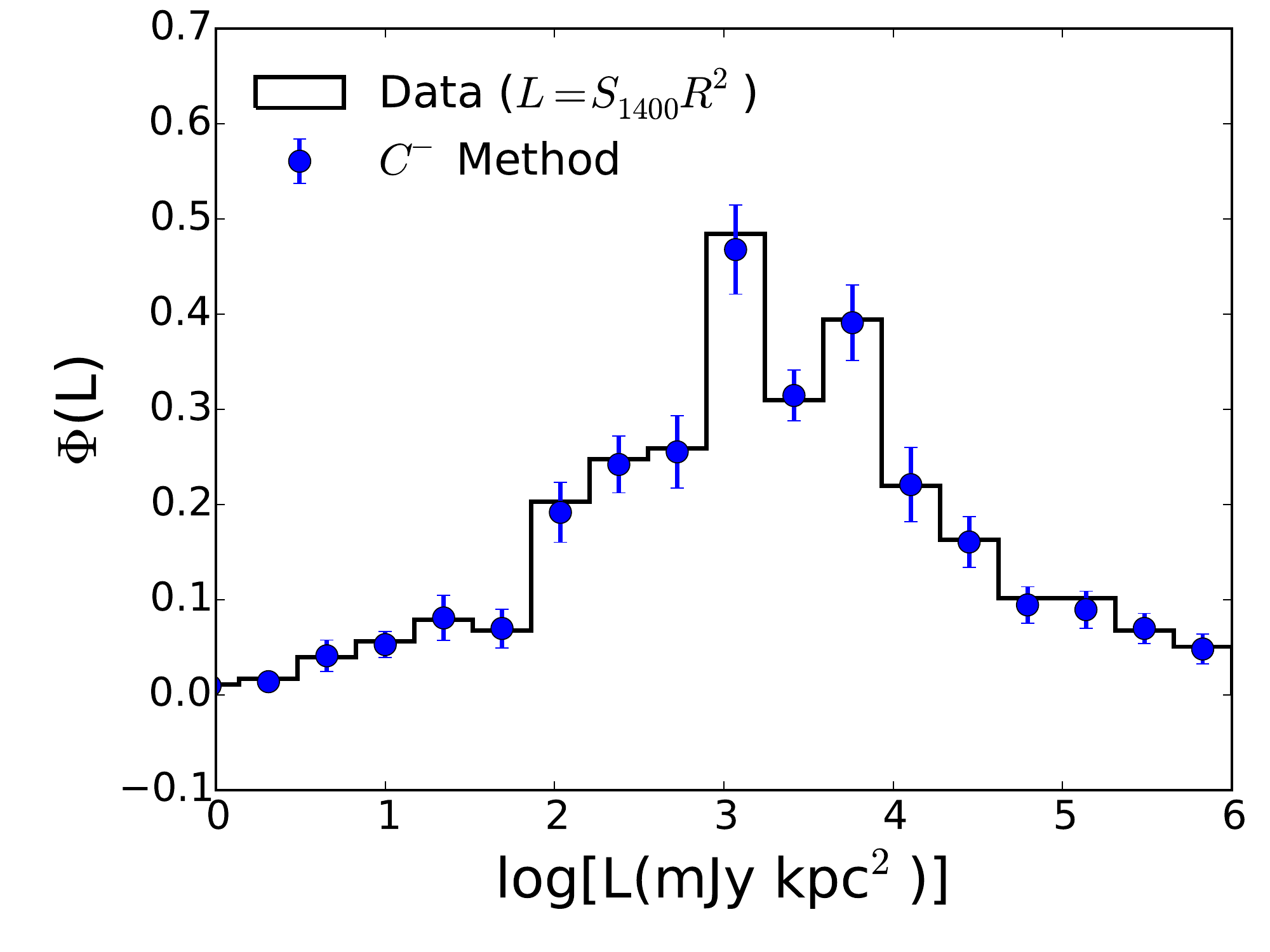}
\includegraphics[width=6cm]{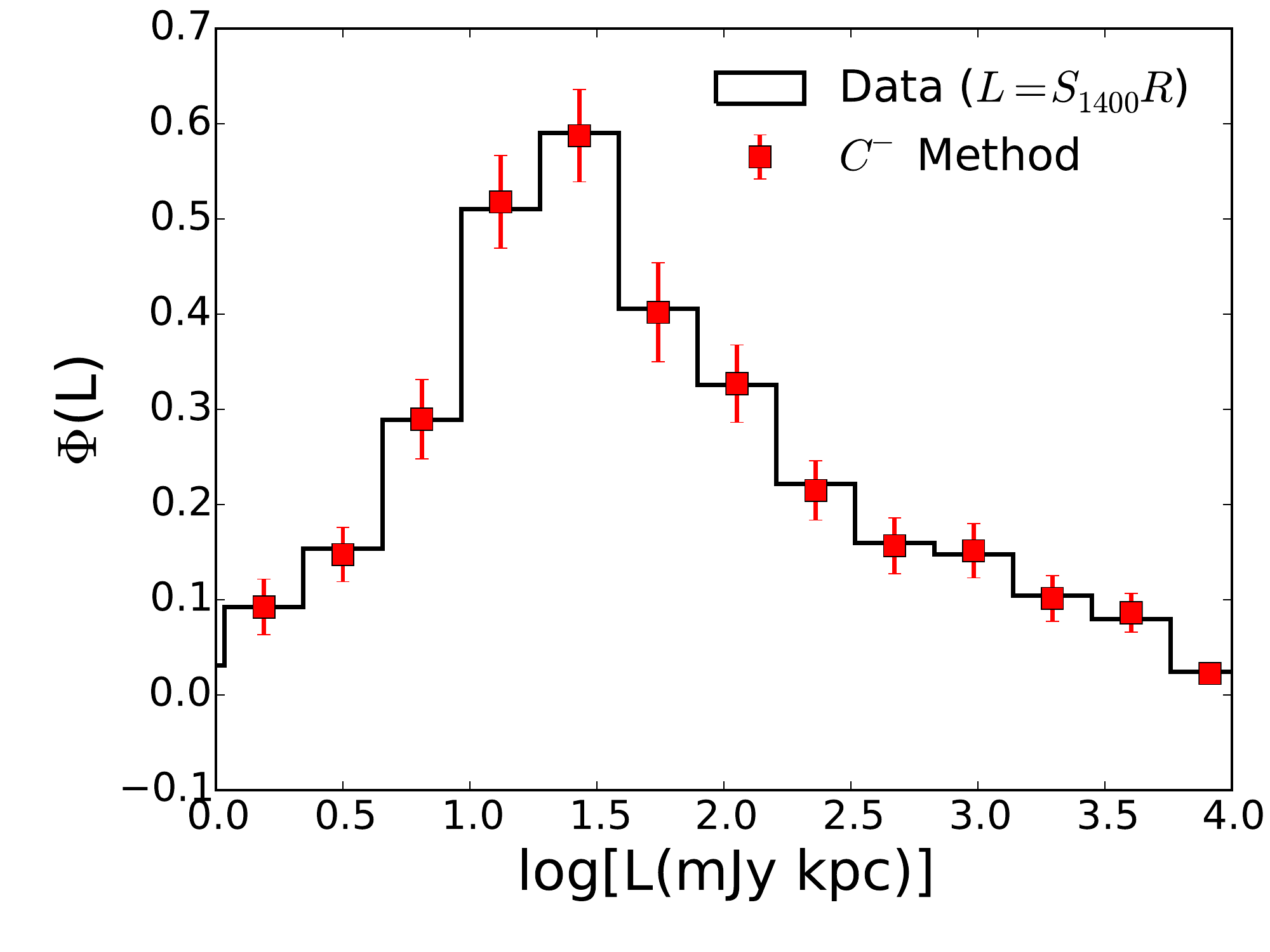}
\caption{Differential normalized luminosity function of pulsars from the Parkes multi-beam survey computed                   
using Lynden-Bell $C^{-}$ method with error bars computed for the usual inverse-square law (top panel) and inverse-linear law (bottom panel). The error bars for both the power-law indices have been estimated using 50 bootstrap resamples. The plots have been made using {\tt astroML}~\citep{astroml}.}
\label{fig:cminus}   
\end{center}
\end{figure}

This  method postulates  that the observed distribution of pulsars  can be derived from a two-dimensional distribution $n(L,r)$ of pulsars with luminosity ($L$) and distance ($r$), where $n(L,r)$ is the probability density function per unit distance and luminosity. It further assumes that the distributions along $L$ and $r$ are uncorrelated  and  the bivariate  distribution can then be separated into functions of luminosity and distance: $n(L,r)=\psi(L) \rho(r)$. The $C^{-}$ method provides a recipe to reconstruct $\psi(L)$ and $\rho(r)$ from the observed dataset. More details  about the implementation of this algorithm  are provided in ~\citet{Lynden-Bell,Jackson,Wilmer,Takeuchi} and we skip the details. Once $\psi(L)$ is estimated, the cumulative luminosity distribution function $\Phi (L)$ can be determined 
as follows: 
\begin{equation}
\phi(L) = \int_{-\infty}^{L} \psi(x) dx
\label{eq:phi}
\end{equation}
The differential luminosity distribution can be obtained by binning  the cumulative luminosity function  obtained from Eq.~\ref{eq:phi}.
We apply the $C^{-}$ method, using the codes provided in  {\tt astroML}~\citep{book} to construct the  differential luminosity function of pulsars 
from Parkes multi-beam survey using the flux at 1400 MHz, after assuming both an  inverse-square law as well as assuming that the flux falls off linearly with distance. For this method, one needs to know the maximum detectable luminosity for the observed distance and the  maximum possible distance to which a pulsar with the observed luminosity can be detected. The former can be estimated by assuming that the maximum distance to which we can detect a pulsar is 20 kpc, and the latter is obtained by assuming that the minimum detectable flux is 0.4 mJy. Therefore, the maximum distance corresponding to a given observed flux ($S$), distance ($r$) and exponent ($n$) is given by $r (S/0.4)^{1/n}$. The differential luminosity distribution of pulsars for both the power law exponents is shown in Fig.~\ref{fig:cminus}. The error bars in each bin for both the exponents are obtained by 50 bootstrap resamples. As we can see from a simple chi-by-eye, the $C^{-}$ method has no problem in reproducing the observed distribution for both the exponents, and the estimated luminosity function is consistent within $1\sigma$ of the observed distribution.  Therefore, although the $C^{-}$ method does not help us to distinguish between the two scenarios, this is the first application of this method in estimating the luminosity function of pulsars.  In future work, we shall also compare the cumulative luminosity function of pulsars using the $C^{-}$ method with other estimates of the same in literature.

\section{Parameter estimation of the distance exponent}

We now address the following question: Can we estimate the distance exponent from the observed pulsar flux with a parameter estimation technique using some variant of the maximum likelihood analysis? This might seem a daunting task, since the physics of pulsar radio emission is not understood~\citep{Hankins}. Furthermore, there is  a large diversity in the observed pulsar population. For example, ~\citet{kjlee} have used machine learning techniques to classify the  pulsar population into five different categories. The emission mechanisms could be different among the disparate pulsar categories. Moreover, the pulsar flux could also depend on parameters which are not always known or easy to estimate (for example, the beaming fraction, radius, mass, or the neutron star equation of state).  However, if the model proposed in ~\citet{Ardavan} to explain the pulsar radio emission is correct, then  pulsars are expected to be standard candles, and we should be able to recover a distance exponent of one
with a likelihood-based regression method. 

Ever since the first pulsar discoveries, a large number of authors have studied how the  pulsar luminosity scales  with  the  pulsar period and its derivative using a variety of  datasets~\citep{Gunn,Narayan81,Narayan87,Hui,Bagchi}. All these fits have been done assuming that the pulsar fluxes obey the inverse-square law. The best-fit values for the various exponents have differed a lot among the authors ~\citep{Bagchi}. In fact,~\citet{Lorimer93} have argued that none of the  proposed scaling laws of pulsar luminosity  with period and its derivative can accurately describe the full pulsar population. 
Nevertheless,  we now generalize this fitting procedure by keeping the distance exponent as a free parameter and use robust methods to see if  we get a good fit, and if the best-fit value of the distance exponent is close to one to vindicate the pulsar emission model of~\citet{Ardavan}.

We model the pulsar flux  as power-law  functions of the distance, the pulsar period and its derivative, and  a normalization constant, and estimate the best-fit values of each of these. This can be thought of as a regression problem to find a relation between the dependent variable (pulsar flux) vs three independent variables: distance, pulsar period, and its first derivative. 
 We now discuss the estimation of the distance exponent from the observed data.   The pulsar flux at 1400 MHz ($S_{1400}$) can be expressed as follows: 
\begin{equation}
S_{1400} = A R^{-n} P^{-q} \dot{P_1}^{m},
\label{eq:flux} 
\end{equation}

\noindent where $\dot{P_1}=10^{15}\dot{P}$,  $R$ is the distance to the pulsar, $P$ is its period, $\dot{P}$ its period derivative,  and $A$ is a normalization constant. We now 
obtain the best-fit estimates of $A$, $n$, $q$, and $m$ using Bayesian statistical inference. The first step in parameter estimation involves constructing the data likelihood ${\cal L}$ given the model and the errors in the data. We use the same data from the Parkes multi-beam survey as S09. We need to  take into account the errors in $S_{1400}$, $R$, and $\dot{P}$. The errors in $S_{1400}$ and $\dot{P}$ are obtained from the online ATNF catalog. For the errors in distance, we use the fractional distance errors from the NE2001 model  as a function of galactic longitude from Fig. 12 of  ~\citet{Lazio}, provided to us by J. Cordes (private communication). Since there are no estimates for these errors  as a function of galactic latitude, we only use the  functional dependence on  galactic longitude to estimate the error in distance for each pulsar. The median errors in distance from the NE2001 model are about 20\%. Recently, ~\citet{Deller} have compared these dispersion based distances  with  parallax based distances and pointed out that the 20\% errors estimated for  the  distance are not realistic. They argue that the distribution of errors cannot be approximated by a single Gaussian, because of a long tail of errors with incorrect distance estimates by more than a  factor of three. (See Fig.~12 of ~\citet{Deller}.) Nevertheless, since we do not have parallax measurements for the full Parkes pulsar sample, we use the distances and the estimated errors from the NE2001  model in this work.
We neglect the errors in $P$, since they are very small compared to the measured values.  We  assume that there  are no covariances in the measured errors between the  different pulsars.

 A number of methods have been developed to incorporate errors in variables on  both sides of Eq.~\ref{eq:flux}, while constructing the likelihood. Here, we follow the formalism of ~\citet{Weiner06}. See also~\citet{Hoekstra}, whose notation we follow. Our likelihood ${\cal L}$ is  as follows:

\begin{equation}
{\cal L}=\prod_{i=1}^{n}\frac{1}{w_i}\exp\left\{-\frac{|S_i-S(P,\dot{P_1},R)|}{2 w_i}\right\}.
\label{eq:likelihood}
\end{equation}
\noindent In Eq.~\ref{eq:likelihood}, $w_i$ for the $i^{th}$ pulsar is given by:
\begin{equation}
w_i^2=\left[\frac{\partial S}{\partial R}\right]^2\sigma_{R_i}^2+\left[\frac{\partial S}{\partial \dot{P}}\right]^2\sigma_{\dot{P}_i}^2+ \sigma_{S_i}^2.
\label{eq:error}
\end{equation}

In Eq.~\ref{eq:error}, ${S_i}$ is the measured flux (at 1400 MHz) for each pulsar, $\sigma_{S_i}$ is the error in the measured flux, $\sigma_{R_i}$ is the error in the measured distance, $\sigma_{\dot{P}_i}$ is the measured error in $\dot{P_1}$. Note that we do not bin the data.  Unlike most literature on parameter estimation we choose a likelihood, which is similar to L1 norm,  or sometimes called M-estimate~\citep{NR}, instead of the widely used L2 norm (or usually known as $\chi^2$/least-squares minimization).  This is to suppress the contribution from outliers and to account for  any diversity in the pulsar population. Note however that our likelihood  differs  from  the M-estimate defined in ~\citet{NR} by an extra  $w_i$ term in the denominator, since 
the error term ($w_i$) is a function of both the dependent and independent variables.  
Alternately, one could also try to bifurcate the pulsar population into a subset of ``standard'' pulsars which obey the scaling relation in Eq.~\ref{eq:likelihood}  and an outlier population, using the methods described in  ~\citet{Hogg} to reject outliers. Although we tried such a  procedure, it is not computationally feasible with an unbinned analysis due to the large number of pulsars.

Once we construct the likelihood, we then calculate the model posterior using Bayes theorem after multiplying the likelihood by a Bayesian prior for each of the unknown parameters:
\begin{equation}
P(M,\theta|D) \propto {\cal L} P(n) P(q) P(m) P(A),
\label{eq:post}
\end{equation}

\noindent where $P(M,\theta|D)$ is the model posterior for the model $M$ given the data $D$, the likelihood (${\cal L}$) is defined in Eq.~\ref{eq:likelihood}, and the vector of parameters $\theta=\{A,m,n,q\}$. $P(n)$, $P(q)$, $P(m)$, and $P(A)$ represent the priors in $n$, $q$, $m$, and $A$ respectively. The Bayesian posterior mean for a given variable $\hat{\theta}$
is given by $\hat{\theta} = \int \hat{\theta} P(\hat{\theta}|D) d\hat{\theta} $, where $P(\hat{\theta}|D)$ for a given
parameter ($\hat{\theta}$) is obtained by marginalizing Eq.~\ref{eq:post} over the other parameters.
 In practice, Eq.~\ref{eq:post} also needs to be normalized by $P(D)$, where $P(D)$ is the probability of the data. However, since we are not doing a model comparison between two distinct sets of models, we shall ignore the normalization term. We choose uniform priors on the parameters with $n \in [-2,20]$, $m \in [-2,20]$,  $q \in [-2,20]$, $A \in [1,1000]$. Using this choice of priors, Eq.~\ref{eq:post} is equivalent to ordinary maximum likelihood analysis. Similar to S09, we only consider pulsars with $S_{1400}>0.4~\mathrm{mJy}$ and distance less than 20 kpc.  With these priors for the four unknown parameters, the best-fit value for the distance exponent $n$ can be found by maximizing  Eq.~\ref{eq:post}, after marginalizing over the other nuisance parameters.

We use the publicly available Markov-chain Monte-Carlo code sampler {\tt emcee}~\citep{emcee} to sample the posterior distribution in  Eq.~\ref{eq:post} and estimate the marginalized best-fit parameters. We start the chain with about 300 `walkers', each of which starts at a different position in parameter space. We run the MCMC for 5000 steps and choose a burn-in of 3000 steps. So the best-fit values are obtained from the last 2000 steps. The best-fit marginalized parameters on each of the four parameters are: $n=1.95 \pm 0.06$, $q=0.41 \pm 0.05$, $m=0.30 \pm 0.01$, and $A=19.5 \pm 2.3$

 It is not straightforward to formally assess the goodness of fit for the Bayesian analysis we have done using a M-estimate based likelihood with an unbinned analysis~\citep{Raja,Lucy}. However, we still need to check whether the best-fit parameters obtained
from our regression analysis provide a good description of the observed fluxes.  From Eq.~\ref{eq:likelihood} and the total number of degrees of freedom (DOF), we calculate -2ln${\cal L}$/DOF (analogous to $\chi^2$/DOF) using our best-fit parameters to see if we get a value close to one.  At the best-fit point, the  value of -2ln${\cal L}$/DOF is about 3.42. Therefore, this is a poor description of the observed fluxes. We also confirmed that the normalized residuals given by  $\frac{S_i-S(P,\dot{P_1},R)}{w_i}$, do not show
a Gaussian distribution with mean at zero. Therefore, we find  (in agreement  with ~\citet{Lorimer93}) that the  pulsar population from the Parkes multi-beam survey  cannot be accurately modeled as  power-law functions of the pulsar period and derivative, after keeping the distance exponent as a free parameter. Moreover, we are unable to  obtain a best-fit value of the distance exponent of one.

Even though we cannot get a good fit to the observed pulsar fluxes,  we  would like to do a model comparison by comparing our best fit with some other distance exponents if they are favored compared to an inverse-square law. To do this, we fix the values of $n$ in Eq.~\ref{eq:post} to 1 and 1.5, and  calculate best-fit values of $m$, $n$, and $A$ for these exponents, and compare the residuals  with our best-fit exponent of 1.95.
This comparison  is shown in Table~\ref{tab:comparechi}. For each value of $n$, we   then calculate  -2ln${\cal L}$/DOF using ${\cal L}$ from Eq.~\ref{eq:likelihood}.  The model with a larger value of -2ln${\cal L}$/DOF is disfavored compared to the  one with  a smaller value. From Table.~\ref{tab:comparechi}, we see that the value for $n=1$ or $n=1.5$ is larger than  for our best-fit value (close to 2), and is therefore disfavored compared to an inverse-square law scaling.

We note that we also tried other variants of the  likelihood besides the one used in  Eq.~\ref{eq:likelihood}, including a binned maximum likelihood analysis. However, none of them provide a good fit to the observed data or recover a best-fit exponent of $n=1$. Therefore, we do not find any evidence
from our likelihood-based analysis that the model for pulsar emission proposed by \cite{Ardavan} can account for the flux of pulsars
from the Parkes multi-beam survey.

\begin{table*}
\small
\caption{Comparison of -2ln${\cal L}$/DOF, where ${\cal L}$ is given by Eq.~\ref{eq:likelihood} and DOF is the total number of degrees of freedom equal to 543, for different values of distance exponent after marginalizing over the nuisance parameters.} 
\label{tab:comparechi}
\begin{tabular}{@{}cr@{}}
\tableline  Exponent & -2ln${\cal L}$/DOF \\
\tableline
1 & 4.05 \\
1.5 & 3.17  \\
1.95 & 3.00 \\
\tableline
\end{tabular}
\end{table*}

\section{Conclusions}
S09~\citep{Singleton} have argued that the radio fluxes of pulsars measured at 1400 MHz from the Parkes multi-beam survey show a violation of the universally accepted inverse-square law behavior. They tried to construct the luminosity function of pulsars (after assuming different power-law dependencies) using the SWML algorithm~\citep{Ellis}.  SWML is an iterative procedure, where one starts with an initial estimate for the luminosity function and the final luminosity function is obtained
after a finite  number of iterations using a bootstrapping procedure. S09 found that the convergence error between successive iterations is smaller for $n=1$ compared to the inverse-square law ($n=2$) by a factor of $10^5$. In this paper, we have tried to verify if the pulsar flux scales inversely with the first power of distance using three different methods. First, we follow exactly the same procedure as S09 and apply the original SWML algorithm to the same pulsar dataset. We posit two different estimates for the luminosity function in the zeroth iteration. The first method assumes that  the luminosity function is inversely proportional to the luminosity, and the second method
uses the observed data  to construct the luminosity function. The final convergence errors for $n=2$ using both these initial guesses   do not agree with the results of S09. Using the first method, we find that the convergence error for $n=2$ is of the same order of magnitude as $n=1$. Using the second method, we find that the convergence error for $n=2$ is only larger by a factor of 10 compared to $n=1$. Therefore, the convergence
error of the SWML algorithm for the pulsar dataset is sensitive to the initial luminosity function used in the zeroth iterations. We are unable to reproduce the results of S09 with two different trial luminosity functions in the zeroth iteration.

We then reconstruct the luminosity function for  the same set of pulsars using the Lynden-Bell $C^{-}$ method~\citep{Lynden-Bell}, after assuming both $n=1$ and $n=2$. We find that in both the cases, the $C^{-}$ method has no problem in reconstructing an empirical luminosity distribution. So this method cannot be used to distinguish between the two distance exponents.

Finally, we extract the distance exponent with a Bayesian regression procedure, after modeling the observed flux  as  power-law functions  of the observed distance, pulsar period,  and the period derivative. We do not get a best-fit value of $n=1$. The best-fit solution we obtain from our regression method   cannot adequately  account for the flux distribution of all the pulsars,
which implies that the flux is also a function of other parameters not accounted for  in our fitting procedure. However, the residuals from our fit for $n=1$ are larger compared to those for $n=2$.

Therefore, using three independent methods we do not find any evidence to support  the claims of S09 that the pulsar flux violates the inverse-square law and the flux decreases linearly with distance.

\section*{Acknowledgments}
We would like to thank John Singleton for a stimulating talk at the 2011 AAS meeting, which provided the impetus for this work. We are grateful to Chris Willmer for explaining the usage of SWML  in extragalactic astronomy and also
to Jim Cordes for providing  us the data for the fractional distance errors as a function of galactic longitude from the NE2001 model. We would like to thank  I-Non Chiu, Alec Habig,  Krishnamoorthy Iyer, and the anonymous referee  for critical feedback on the paper draft. 

\nocite{*}
\bibliographystyle{spr-mp-nameyear-cnd}

\end{document}